\documentclass[preprint,prc,showpacs,preprintnumbers,
               superscriptaddress,floatfix]{revtex4}

\usepackage{graphicx,amsmath,amssymb}
\usepackage{dcolumn}
\usepackage{bm}
\usepackage{longtable}

\newcommand{\beq}{\begin{equation}}
\newcommand{\eeq}{\end{equation}}
\newcommand{\beqn}{\begin{eqnarray}}
\newcommand{\eeqn}{\end{eqnarray}}
\newcommand{\btab}{\begin{tabular}}
\newcommand{\etab}{\end{tabular}}

\begin{document}

\title{Microscopic linear response calculations based on the \\
      Skyrme functional plus the pairing contribution }

\author{ Jun Li 
}
\thanks{ jun.li@mi.infn.it }
\affiliation{ Dipartimento di Fisica, Universit\`a degli Studi
              and INFN Sez. di Milano, Via Celoria 16, 20133, Milano, Italy }
\affiliation{ School of Physics, and SK Lab. Nucl. Phys. $\&$ Tech.,
              Peking University, Beijing 100871, China }

\author{ Gianluca Col\`o }
\thanks{ gianluca.colo@mi.infn.it }
\affiliation{ Dipartimento di Fisica, Universit\`a degli Studi
              and INFN Sez. di Milano, Via Celoria 16, 20133, Milano, Italy }

\author{ Jie Meng 
}
\thanks{ mengj@pku.edu.cn }
\affiliation{ School of Physics, and SK Lab. Nucl. Phys. $\&$ Tech.,
              Peking University, Beijing 100871, China }
\affiliation{Department of Physics, University of Stellenbosch,
             Stellenbosch, South Africa}
\affiliation{ Institute of Theoretical Physics, Chinese Academy of
              Science, Beijing, 100080, China }
\affiliation{ Center of Theoretical Nuclear Physics, National
              Laboratory of Heavy Ion Accelerator, Lanzhou, 730000, China }


\begin{abstract}

A self-consistent Quasiparticle-Random-Phase-Approximation (QRPA) model which
employs the canonical Hartree-Fock-Bogoliubov (HFB) basis and an energy-density
functional with a Skyrme mean field part and a density-dependent pairing, is
used to study the monopole collective excitations of spherical even-even nuclei.
The influence of the spurious state on the strength function of the isoscalar
monopole excitations is clearly assessed. We compare the effect of different kinds
of pairing forces (volume pairing, surface pairing and mixed pairing) on the
monopole excitation strength function. The energy of the Isoscalar Giant Monopole
Resonance (ISGMR), which is related to the nuclear incompressibility $K_{\infty}$,
is calculated for tin isotopes and the results are discussed.

\end{abstract}


\maketitle

\section {introduction}

The nuclear structure community have made great achievements in understanding
the structure of the ground state and of the excited states of stable atomic nuclei.
As radioactive beams provide more experimental results on the nuclei far from the
stability valley, the challenge is how to extrapolate the theoretical models and
predict or describe in detail the exotic properties of the nuclei with large neutron
or proton excess. Another challenge is the prediction of the properties of nuclear matter
in a broad range of densities, i.e., in connection with neutron stars, and to
understand the origin of these new properties.

For medium-mass and heavy nuclei, the most microscopic models that we can use are
the mean field models based on the effective interactions, either in the
non-relativistic or relativistic framework. For closed shell nuclei, Hartree-Fock (HF)
theory has already been proven to be a powerful tool to describe the properties of
ground states \cite{ring1980}, in particular using the zero-range Skyrme interactions
\cite{vautherin1972,beiner1975,sg2,skm*,sly5}. In the open shell nuclei, the effect of
nuclear pairing shows up. A simple theory for the ground-state pairing is HF+Bardeen-Cooper-Schriffer
(BCS)\cite{cooper1957pr}. The nuclei close to the neutron or proton drip lines, may
exhibit some very unusual features such as the neutron or proton skin \cite{fukunishi1993}
and the neutron haloes \cite{hansen1995}. In these very neutron-rich or proton-rich nuclei,
nuclear pairing plays a crucial role for the theoretical understanding of these new
phenomena \cite{bertsch1991,bertulani1993,meng1996,meng2006,barranco2001}. A more
appropriate theory is the HFB approach \cite{ring1980} because the pairing component can no
longer be treated as a residual interaction, i.e., a small perturbation important only
in the neighborhood of the Fermi surface, as in the nuclei close to the line of beta
stability. This is seen from the approximate HFB relations between the Fermi level
$\lambda$, pairing gap $\Delta$, and the particle separation energy
$s$: $s\approx-\lambda-\Delta$, since $s$ is very small and $\lambda+\Delta\approx0$,
for drip-line nuclei. Consequently, the mean field characterized by $\lambda$ and the
pairing field $\Delta$ are equally important. Using appropriate effective interactions in
mean field and pairing field, the HFB approach is already sophisticated enough to allow
precise analysis of ground state properties, i.e., binding energies, average neutron
pairing gaps, etc., in most nuclei either using the Skyrme force plus a density-dependent
pairing force \cite{nazarewicz1994} or the finite-range pairing
force \cite{decharge1980,berger1991}.

Studying the nuclear collective excitations is another important tool to understand
the structure of nuclei and predict the exotic properties of nuclei far from
stability valley or the properties of nuclear matter. The QRPA is a standard method
for describing these collective excitations in open-shell superconducting nuclei with
stable mean field solutions \cite{ring1980,rowe1970}. Important nuclear collective
excitations are the nuclear compressional modes -- particularly the ISGMR -- which
provide the optimal route to determine the nuclear incompressibility
\cite{colo2004,shlomo2006,colo2008}. Both non-relativistic RPA \cite{peru2005,sil2006,sagawa2007}
and relativistic RPA or QRPA \cite{vretenar2003,piekarewicz2007} were recently used
in studying the nuclear collective excitations and the nuclear incompressibility.
However, the fully self-consistent QRPA, formulated in the HFB canonical basis,
which was introduced and accurately tested using Skyrme energy density functionals
and density-dependent pairing functionals in Ref. \cite{terasaki2005} has not been
applied to extract information about the nuclear incompressibility.
If the models are characterized by a nuclear incompressibility
$K_{\infty}$ around 230$\sim$240 MeV (or 250$\sim$270 MeV in the case of relativistic
mean field (RMF)), they will give the right ISGMR centroid energies compared with the
experimental data in $^{208}$Pb \cite{vretenar2003,colo2004}. However, with these modes
one tends to overestimate the centroid energies in the tin isotopes
\cite{piekarewicz2007,sagawa2007}. Is there any need of more sophisticated considerations
for the nuclear pairing in tin ? To answer this question, we study the ground state
properties of tin isotopes using the HFB theory with a Skyrme mean field and different
kind of pairing forces, and then we apply a self-consistent QRPA model for the ISGMR.

The paper is organized as follows. Section II discusses the QRPA employed in this study.
Since the QRPA equation is solved on the canonical HFB basis, in this section we also
present a survey of the canonical HFB theory. We mention briefly the
meaning and role of self-consistency in the context of RPA and QRPA calculations. In section III, we give the numerical
details and the results. In this section, we discuss the method of projecting out
the spurious state, the influence of pairing force on the strength function, and
in particular, we study the energy of the ISGMR for tin isotopes. The last section
contains a short summary and our conclusions.

\section{QRPA with canonical HFB basis}

The standard approach for describing collective excitations in magic nuclei is
the so-called Random-Phase-Approximation (RPA) with stable mean-field solutions.
The extension to open-shell nuclei is the QRPA which includes not only the
particle-hole channel (mean field), but also the particle-particle channel
(pairing field). The first step in solving the self-consistent QRPA is to obtain
the ground state properties by solving the HFB equations. Since the HFB theory
is discussed in details in many textbooks and articles \cite{ring1980,nazarewicz1994},
here we just present briefly the HFB theory in canonical basis.

The canonical basis is a particular basis in which the density matrix is diagonal,
and the canonical basis wave functions are all localized. Starting from the
effective Hamiltonian of a nuclear system, using the variational principle,
the HFB equations in the canonical basis are obtained as \cite{nazarewicz1994}
\beqn
\label{canhfb1}
   (h - \lambda)_{\mu\nu}(u_{\mu}v_{\nu} + u_{\nu}v_{\mu})
   + \tilde{h}_{\mu\nu}(u_{\mu}u_{\nu} - v_{\nu}v_{\mu})
&=&
   0
\\
\label{canhfb2}
   (h - \lambda)_{\mu\nu}(u_{\mu}u_{\nu} - v_{\nu}v_{\mu})
   - \tilde{h}_{\mu\nu}(u_{\mu}v_{\nu} + u_{\nu}v_{\mu})
&=&
   E_{\mu\nu},
\eeqn
where $h$ and $\tilde{h}$ are the particle-hole and particle-particle
Hamiltonian in the canonical basis, respectively. $\lambda$ is the Fermi
energy while $E_{\mu\nu}$ represents the HFB Hamiltonian in
the canonical basis. The canonical
states $\mu$ are not eigenstates of the HFB Hamiltonian. The right way to obtain
the canonical basis is to solve the HFB equations in the quasiparticle basis to
obtain the quasiparticle states, i.e., the quasiparticle wave functions (including
the upper and lower conponents $\phi_{1}$ and $\phi_{2}$) and quasiparticle energies,
and then to build the density matrix and to diagonalize it
to get the canonical basis
\cite{nazarewicz1994}. The square of $v_{\mu}$ ($u_{\mu}$) represent the
probability that a certain state $\mu$ is occupied (empty). These quantities are
solely determined by the diagonal matrix elements of particle-hole and particle-particle
Hamiltonians,
\beqn
   v_{\mu}
&=&
   -\textrm{sgn}(\tilde{h}_{\mu\mu})\sqrt{\frac{1}{2}
   - \frac{h_{\mu\mu}-\lambda}{2E_{\mu\mu}}},
~~~~~
   u_{\mu}
~=~
   \sqrt{\frac{1}{2} + \frac{h_{\mu\mu}-\lambda}{2E_{\mu\mu}}}.
\eeqn

In the case of spherical symmetry, and without mixing of the proton
and neutron states,
the quasiparticle wave functions have the good quantum numbers ($nljm$);
$n$ turns out to be a good quantum number because the continuum
is discretized inside a
spherical box. Furthermore, the radial part of the quasiparticle wave functions
can be chosen to be real. Then the quasiparticle wave function can be written as:
\beqn
   \phi_{i}(nlj,\mathbf{r}\sigma)
&=&
   \frac{u_{i}(nlj,r)}{r}\sum_{m_{l}m_{s}
   }Y_{m_{l}}^{(l)}(\hat{r})\chi_{m_{s}}^{(\sigma)}\langle lm_{l}\frac{1}{2}m_{s}|jm\rangle,
   ~~~~i=1,2,
\eeqn
where $i=1,2$ label the upper and lower components of the wave functions.
The density matrix elements can be constructed using the lower components of
the quasiparticle wave functions,
\beqn
   \rho_{(lj)}(\textbf{r},\textbf{r}^{\prime})
&=&
   \sum_{n}\phi_{2}(nlj,\textbf{r})\phi^{\ast}_{2}(nlj,\textbf{r}^{\prime}),
\eeqn
where the sum runs over all the states for a given $(l,j)$ block.
We will get the canonical basis after diagonalizing the density matrix.

The QRPA takes the quasiparticle vacuum as the approximate ground state and
is aimed at the description of small amplitude, collective excitations.
Defining the phonon excitation operators in the angular momentum coupled
representation,
\beqn
   Q_{\nu}^{\dag}(JM)
&=&
   \sum_{\alpha\geq\beta}(X_{\alpha\beta}^{\nu}A_{\alpha\beta}^{\dag}(JM)
   - Y_{\alpha\beta}^{\nu}A_{\alpha\beta}(\widetilde{JM}))
\\
\label{qnu}
   Q_{\nu}(JM)
&=&
   \sum_{\alpha\geq\beta}(X_{\alpha\beta}^{\nu\ast}A_{\alpha\beta}(JM)
   - Y_{\alpha\beta}^{\nu\ast}A_{\alpha\beta}^{\dag}(\widetilde{JM})),
\eeqn
where
\beqn
   A_{\alpha\beta}^{\dag}(JM)
&=&
   \frac{1}{\sqrt{1+\delta_{\alpha\beta}}}\sum_{m_{\alpha}m_{\beta}}
   \langle j_{\alpha}m_{\alpha}j_{\beta}m_{\beta}|JM\rangle
   \alpha_{\alpha}^{\dag}\alpha_{\beta}^{\dag}
\\
   A_{\alpha\beta}(JM)
&=&
   \frac{1}{\sqrt{1+\delta_{\alpha\beta}}}\sum_{m_{\alpha}m_{\beta}}
   \langle j_{\alpha}m_{\alpha}j_{\beta}m_{\beta}|JM\rangle
   \alpha_{\beta}\alpha_{\alpha}
\\
   A_{\alpha\beta}(\widetilde{JM})
&=&
   (-1)^{J+M}A_{\alpha\beta}(J-M),
\eeqn
the QRPA equation can be easily obtained from the linearization of Schr\"{o}dinger equation,
\beqn
   \sum_{\alpha<\beta}
   \left(
      \begin{array}{c c}
         A_{\alpha\beta,\gamma\delta}         & B_{\alpha\beta,\gamma\delta}         \\
         -B^{\ast}_{\alpha\beta,\gamma\delta} & -A^{\ast}_{\alpha\beta,\gamma\delta} \\
      \end{array}
   \right)
   \left(
      \begin{array}{c}
         X^{\nu}_{\gamma\delta}  \\ Y^{\nu}_{\gamma\delta} \\
      \end{array}
   \right)
&=&
   E_{\nu}
   \left(
      \begin{array}{c}
         X^{\nu}_{\alpha\beta}  \\ Y^{\nu}_{\alpha\beta} \\
      \end{array}
   \right).
\eeqn

In the angular momentum coupled representation and canonical HFB basis,
the matrix elements $A_{\alpha\beta,\gamma\delta}$ and $B_{\alpha\beta,\gamma\delta}$
have the form \cite{ring1980,terasaki2005,rowe1970},
\beqn
    A_{\alpha\beta,\gamma\delta}
&=&
   \frac{1}{\sqrt{1+\delta_{\alpha\beta}}\sqrt{1+\delta_{\gamma\delta}}}
\nonumber\\
&&\times
   \left[
      E_{\alpha\gamma}\delta_{\beta\delta} -
      (-1)^{j_{\alpha}+j_{\beta}-J}E_{\beta\gamma}\delta_{\alpha\delta}
      - (-1)^{j_{\alpha}+j_{\beta}-J}E_{\alpha\delta}\delta_{\beta\gamma}
      + E_{\beta\delta}\delta_{\alpha\gamma}
   \right.
\nonumber\\
&&
      + G_{\alpha\beta\gamma\delta}
      (u_{\alpha}u_{\beta}u_{\gamma}u_{\delta} + v_{\alpha}v_{\beta}v_{\gamma}v_{\delta})
      + F_{\alpha\beta\gamma\delta}
      (u_{\alpha}v_{\beta}u_{\gamma}v_{\delta} + v_{\alpha}u_{\beta}v_{\gamma}u_{\delta})
\nonumber\\
&&
   \left.
      - (-1)^{j_{\gamma}+j_{\delta}-J^{\prime}}F_{\alpha\beta\delta\gamma}
      (u_{\alpha}v_{\beta}v_{\gamma}u_{\delta} + v_{\alpha}u_{\beta}u_{\gamma}v_{\delta})
   \right]
\\
   B_{\alpha\beta,\gamma\delta}
&=&
   \frac{1}{\sqrt{1+\delta_{\alpha\beta}}\sqrt{1+\delta_{\gamma\delta}}}
\nonumber\\
&&\times
   \left[
      - G_{\alpha\beta\delta\gamma}
      (u_{\alpha}u_{\beta}v_{\gamma}v_{\delta} + v_{\alpha}v_{\beta}u_{\gamma}u_{\delta})
      - (-1)^{j_{\delta}+j_{\gamma}-J^{\prime}}F_{\alpha\beta\delta\gamma}
      (u_{\alpha}v_{\beta}u_{\gamma}v_{\delta} + v_{\alpha}u_{\beta}v_{\gamma}u_{\delta})
   \right.
\nonumber\\
&&
   \left.
      + (-1)^{j_{\alpha}+j_{\beta}+j_{\gamma}+j_{\delta}-J-J^{\prime}}F_{\alpha\beta\gamma\delta}
      (u_{\alpha}v_{\beta}v_{\gamma}u_{\delta} + v_{\alpha}u_{\beta}u_{\gamma}v_{\delta})
   \right]
\eeqn
with
\beqn
   G_{\alpha\beta\gamma\delta}
&=&
   \sum_{m_{\alpha}m_{\beta}m_{\gamma}m_{\delta}}
   \langle j_{\alpha}m_{\alpha}j_{\beta}m_{\beta}|JM\rangle
   \langle j_{\gamma}m_{\gamma}j_{\delta}m_{\delta}|J^{\prime}M^{\prime}\rangle
   V^{pp}_{\alpha\beta,\gamma\delta}
\\
   F_{\alpha\beta\gamma\delta}
&=&
   \sum_{m_{\alpha}m_{\beta}m_{\gamma}m_{\delta}}
   \langle j_{\alpha}m_{\alpha}j_{\beta}m_{\beta}|JM\rangle
   \langle j_{\gamma}m_{\gamma}j_{\delta}m_{\delta}|J^{\prime}M^{\prime}\rangle
   V^{ph}_{\alpha\bar{\delta}\bar{\beta}\gamma},
\eeqn
where $V^{ph}_{\alpha\bar{\delta}\bar{\beta}\gamma}$ represent the matrix elements
of the particle-hole effective interaction, and $V^{pp}_{\alpha\beta,\gamma\delta}$
represent the matrix elements of the particle-particle effective interaction.
Actually, in the canonical basis, the equations of HFB+QRPA are almost the same as
those of HF-BCS+QRPA, except for the presence of off-diagonal terms
related to the quasiparticle energies in the former case.

We remind the reader that self-consistency means, in the RPA context, that
the particle-hole residual interaction is obtained as the exact second
derivative of the energy functional with respect to the density. Up to
quite recently, most of the RPA calculations were {\em not} fully
self-consistent, rather some parts of the residual interaction were
neglected. Only full self-consistency allows conservation of
the energy-weighted sum rule (EWSR) according to the Thouless
theorem~\cite{thouless1961}. Moreover, in the case of monopole the
inclusion of the whole residual interaction is crucial in order
to evaluate accurately the energy of the giant resonance~\cite{sil2006},
and to assess quantitatively the value of $K_\infty$~\cite{colo2004}.

In the QRPA framework, self-consistency should mean, in addition to
what has been stated for RPA, that the particle-particle residual
interaction is obtained as the second derivative of the energy
functional with respect to the anomalous density or pairing
density. This prescription is obeyed both in HF-BCS+QRPA and in
HFB+QRPA. In the former case, however, the model spaces for
HF-BCS (which includes, as a rule, only one major shell above the
Fermi energy) and for QRPA (which is much larger), differ. In this
sense, only HFB+QRPA is, strictly speaking, self-consistent.
The Thouless theorem concerning the EWSR obtained within HF plus
RPA is also valid for the self-consistent QRPA based on HFB solution
as it is demonstrated in Ref.~\cite{khan2002}.

The fully self-consistent calculation must include the
pairing-rearrangement terms in the matrix elements, namely the
contribution to the particle-hole interaction coming from the density
dependence of the pairing force \cite{terasaki2005,waroquier1987}.
In the case of the volume pairing force, there are consequently no pairing-rearrangement
terms at all, while in the case of the surface pairing force there exists
a contribution from the pairing-rearrangement terms. We have carefully
checked that this contribution
has a negligible effect on the ISGMR. For example, the ISGMR centroid energies
obtained from the QRPA calculations based on canonical HFB solution
without the effects of the pairing-rearrangement using the SKM*
force and the surface pairing force are 16.32, 16.13, and 16.0 MeV
for $^{112}$Sn, $^{116}$Sn and $^{124}$Sn, respectively. When including the
effects of the pairing-rearrangement, the values of the ISGMR centroid energies
change to 16.34, 16.11, and 16.02 MeV for $^{112}$Sn, $^{116}$Sn and $^{124}$Sn,
respectively. Therefore, in the figures and in the discussion below, we
quote results obtained without the pairing-rearrangement terms.

\section{results and discussions}

For the ground state properties of spherical nuclei, the HFB equations
are solved in coordinate space in a spherical box. In the particle-hole channel,
we use a Skyrme force, i.e., SLy5 \cite{sly5} or SKM* \cite{skm*}, and in the
particle-particle channel, we use a zero-range density-dependent pairing force,
\beqn
   v(\textbf{r}_{1},\textbf{r}_{2})
&=&
   v_{0}
   \left[
      1 - \eta
      \left(
         \frac{\rho(\frac{\textbf{r}_{1}+\textbf{r}_{2}}{2})}{\rho_{0}}
      \right)
   \right]\delta(\textbf{r}_{1}-\textbf{r}_{2}),
\eeqn
where $\rho_{0}=0.16$ fm$^{-3}$; the values of $\eta$ are
0, 1, and 0.5 for volume, surface and mixed pairing forces, respectively;
the value of $v_{0}$ is fixed by fitting the experimental data of the mean neutron
gap of $^{120}$Sn ($\Delta_{n}=1.321$ MeV). In the case of the Skyrme force SLy5,
$v_{0}$= -170.92, -537.05 and -272.99 MeV fm$^{3}$ for volume, surface and
mixed pairing forces, respectively. In the case of the Skyrme force SKM*,
$v_{0}$= -142.01, -490.48 and -233.22 MeV fm$^{3}$ for volume, surface and
mixed pairing forces, respectively. There are some numerical parameters in
the actual calculations: (i) We use a high quasiparticle-energy cutoff
(200 MeV) and a maximum angular momentum $j_{max} = 15/2$. (ii) The radius
of the box is fixed at 20 fm with a small mesh (0.05 fm).

\begin{figure}[htbp]
  \centering
  \includegraphics[height=10cm,width=13cm]{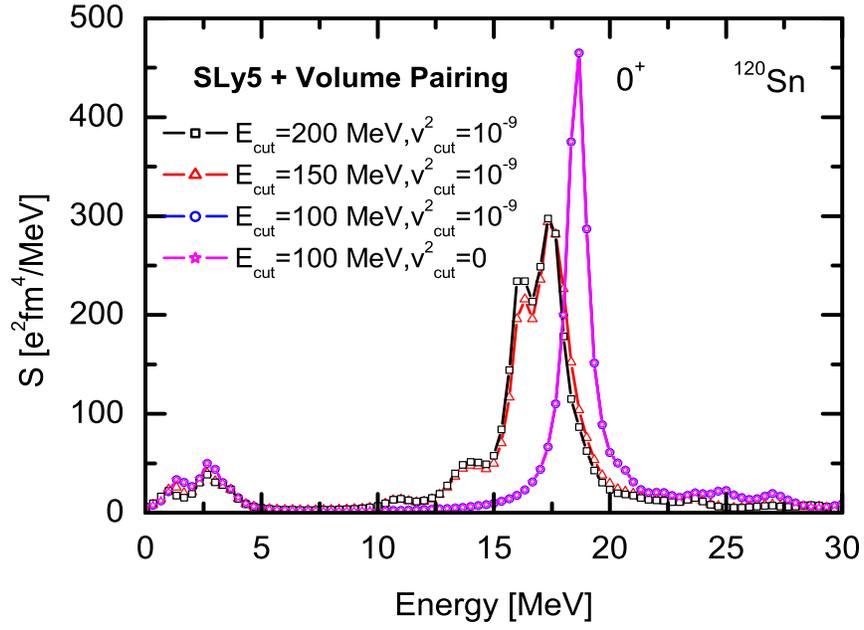}
  \caption{(Color online)The isoscalar $0^{+}$ strength function (averaged with
            Lorentzians having 1 MeV width) in $^{120}$Sn using the Skyrme force SLy5
            and the volume pairing force for different values of single particle energy
            cutoff and occupation probabilities.}
  \label{fig:fig1}
\end{figure}

The self-consistency of the HFB+QRPA calculations requires the use in QRPA
of a residual force derived from the HFB fields. At the same time, for a real
self-consistent calculation, all the quasiparticle states produced by the HFB
calculation must be used to build the matrices A and B in the QRPA equation.
In actual calculation, there are about 500-600 quasiparticle states in the HFB
calculation up to 200 MeV which is the HFB cutoff. Some states with very small
values of occupation probability in canonical basis or some two quasiparticle
excitations with very high energy give little contribution to the QRPA spectrum,
so we cut off the canonical-basis wave functions by excluding those
with very small values
of occupation probability and with
very high values of single particle energy, and
check the convergence of the QRPA solution. Fig. \ref{fig:fig1} shows the isoscalar
$0^{+}$ strength function (averaged with Lorentzians having 1 MeV width) in
$^{120}$Sn with the Skyrme force SLy5 and the volume pairing force defined above,
for different values of cutoff in excitation energy and occupation probability.
Comparing the curve with circles with the curve with stars in Fig. \ref{fig:fig1},
we find that the levels with occupation probabilities smaller than $10^{-9}$ have
little influence on the isoscalar $0^{+}$ strength function.
Similarly, from the curves with squares, up-triangle, and circles in Fig. \ref{fig:fig1},
we find E$_{cut}=150\sim200$ MeV is a suitable excitation energy cutoff in calculations to
make our results stable at the level of 300 keV for the the centroid energy. Therefore,
in the next calculations, the value of the cutoff on occupation probability is fixed
at $v^{2}_{cut} = 10^{-9}$, and the excitation energy cutoff is selected as
E$_{cut} = 200$ MeV.

\subsection{The spurious state}

In the spherical QRPA solution, there exists the problem of a spurious state in
the monopole channel due to the particle number symmetry broken by the HFB solution
\cite{ring1980}. The spurious state should be orthogonal to all other
physical states, and should appear at zero energy. However, in actual calculations
the spurious state is at low (but not zero) energy
for small numerical inaccuracies,
and contributes to the total strength. Therefore, the spurious state must
be projected out from the real physical states.

Starting from the actual quasiparticle QRPA set of states $|n\rangle$, we construct
a new set of normalized states $|n^{\prime}\rangle$ which are the real physical states,
\beqn
   |n^{\prime}\rangle
&=&
   |n\rangle - \alpha_{n}|s\rangle
\eeqn
where $|s\rangle$ is the spurious state.
The $X$- and $Y$-amplitudes of the spurious state should be proportional to \cite{ring1980}
\beqn
   \langle\alpha\beta|N|0\rangle
&=&
   - 2\hat{j}_{\alpha}u_{\alpha}v_{\alpha},
\eeqn
where $|\alpha\beta\rangle$ is a pair of canonical states, \textit{N} is the particle
number operator, $|0\rangle$ is the phonon vacuum state, and $\hat{j}=\sqrt{2j+1}$.
$\alpha_{n}$ is obtained by the condition $\langle n^{\prime}|N|0\rangle=0$,
\beqn
   \alpha_{n}
&=&
   \frac{\langle n|N|0\rangle}{\langle s|N|0\rangle}
~=~
   - \frac
   {\sum_{\alpha}(X_{\alpha\alpha} + Y_{\alpha\alpha})2\hat{j}_{\alpha}u_{\alpha}v_{\alpha}}
   {\sum_{\alpha}(2\hat{j}_{\alpha}u_{\alpha}v_{\alpha})^{2}}.
\eeqn

\begin{figure}[htbp]
  \centering
  \includegraphics[height=10cm,width=13cm]{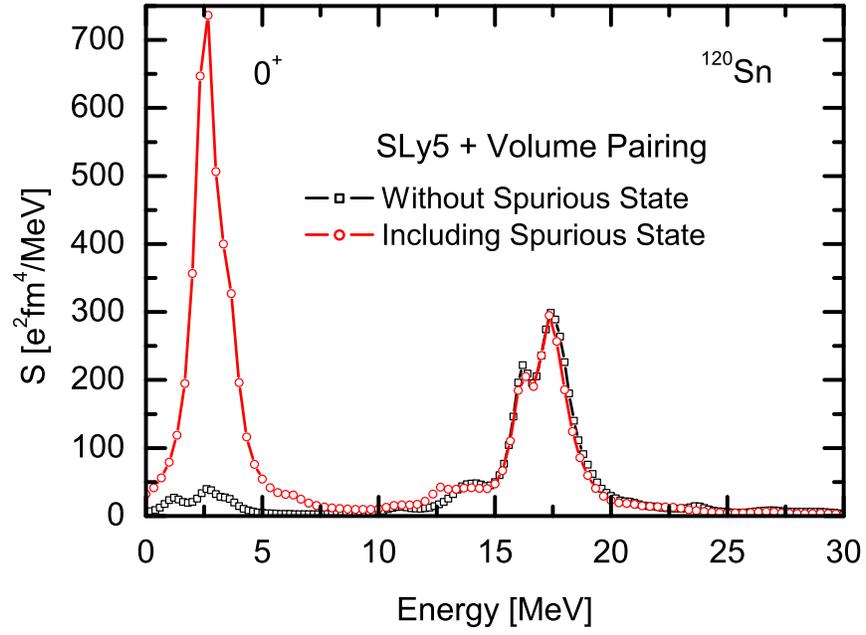}
  \caption{(Color online)The isoscalar $0^{+}$ strength function (averaged with
            Lorentzians having 1 MeV width) including spurious state (line
            with circles) or not (line with squares) in $^{120}$Sn using
            the Skyrme force SLy5 and the volume pairing force.}
  \label{fig:fig2}
\end{figure}

Fig. \ref{fig:fig2} shows the influence of the spurious state on the isoscalar $0^{+}$
strength function in $^{120}$Sn with the Skyrme force SLy5 and the volume pairing force.
The spurious state affects the low-lying region as it is expected, while it does
not impact the ISGMR at all.

\subsection{The isoscalar $0^{+}$ mode}

The ISGMR is an important collective excitation because its excitation
energy is related to the nuclear incompressibility $K_{\infty}$.
The excitation operator of the ISGMR is $F^{IS}_{monopole} = \sum_{i=1}^{A}r_{i}^{2}$.
The macroscopic picture for a state excited by this operator is the so-called
``breathing mode''. The strength function is defined as
\beqn
   S(E)
&=&
   \sum_{n}|\langle n|F^{IS}_{monopole}|0\rangle|^{2}\delta(E-E_{n}),
\eeqn
and the moments of the strength function are
\beqn
   m_{k}
&=&
   \int E^{k}S(E)dE.
\eeqn
The centroid energy can be defined as the ratio between the EWSR sum rule
$m_{1}$ and the non-energy weighted (NEWSR) sum rule $m_{0}$: $E_{0}=m_{1}/m_{0}$.

\begin{figure}[htbp]
  \centering
  \includegraphics[height=10cm,width=13cm]{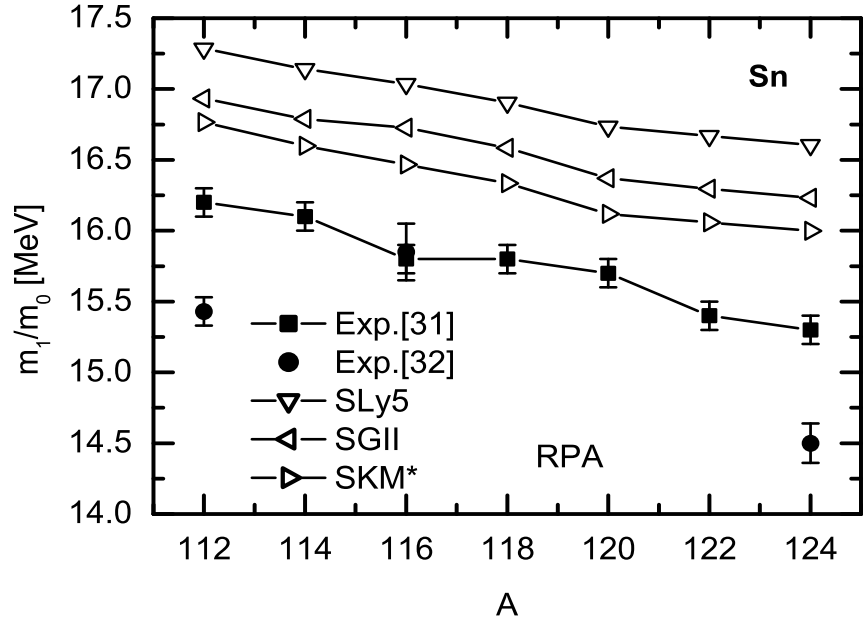}
  \caption{Systematics of the moment ratios $m_{1}/m_{0}$ for the ISGMR
           strength distributions in the tin isotopes using Skyrme HF+RPA.
           They are evaluated in the energy interval between 10.5 and 20.5 MeV.
           The experimental data are extracted from Ref. \cite{lit2007,youngblood2004}}
  \label{fig:fig3}
\end{figure}

\begin{figure}[htbp]
  \centering
  \includegraphics[height=10cm,width=13cm]{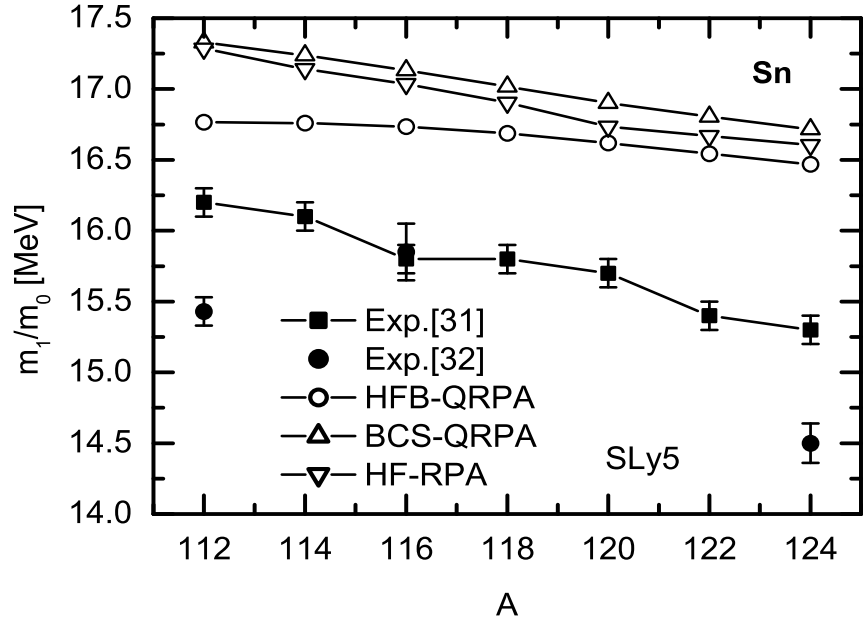}
  \caption{Systematics of the moment ratios $m_{1}/m_{0}$ for the ISGMR
           strength distributions in the tin isotopes.
           They are evaluated in the energy interval between 10.5 and 20.5 MeV.}
  \label{fig:fig4}
\end{figure}

The ISGMR has been so far mainly described within the HF+RPA
and the HF-BCS+QRPA which have been defined above.
In Fig. \ref{fig:fig3} we show the results for the ISGMR centroid energies
in the tin isotopes using HF+RPA (by employing the Skyrme forces
SLy5~\cite{sly5}, SGII~\cite{sg2}, and SKM*~\cite{skm*}). The centroid energies
are evaluated in the energy interval between 10.5 and 20.5 MeV, and the
experimental data come from Ref. \cite{lit2007}. Other
experiments devoted to the ISGMR in tin isotopes have been reported
in Refs.~\cite{youngblood2004}. Their results are also shown,
although the centroid energies are evaluated in the energy interval
between 10 and 35 MeV. In the present work, we have no clear way
to judge the difference between the two experimental results.
The trend along the isotope chain is similar for the three forces. The force
SLy5 ($K_{\infty}$ = 233.8 MeV) gives higher centroid energy than the others
(which have $K_{\infty}$ around 215 MeV). HF+RPA overestimates the centroid
energies by about 1 MeV in all the measured tin isotopes, and there exists a
difference by few hundreds of keV between different Skyrme forces.
The force SLy5 reproduces well the monopole energy in $^{208}$Pb.
The question why it overestimates this energy in the tin isotopes has been raised,
as it is recalled in the introduction. If we take pairing into account, by means
of the HF-BCS+QRPA, the results are very similar to those from HF+RPA, as it is
expected and shown in Fig. \ref{fig:fig4}, which presents the ISGMR centroid
energies in the tin isotopes by using SLy5 and the volume pairing force.

One possible guess could be that theoretical models overestimate the centroid energies in
the tin isotopes because a more appropriate method to deal with the pairing correlations
than HF-BCS+QRPA, i.e., the HFB+QRPA, is needed. As shown in Fig. \ref{fig:fig4},
HFB+QRPA makes the theoretical results significantly closer to experiment, especially
in $^{112,114,116}$Sn. However, this improvement is not enough. The difference between
HF-BCS+QRPA (HF+RPA) and HFB+QRPA comes from the fact that in $^{112,114,116}$Sn,
the single particle energies and occupation number of d3/2 and h11/2 are different
for HF-BCS (HF) and HFB, while they are similar in $^{120,122,124}$Sn. These numbers
are shown in Table \ref{tab:table1}.

\begin{figure}[htbp]
  \centering
  \includegraphics[height=10cm,width=13cm]{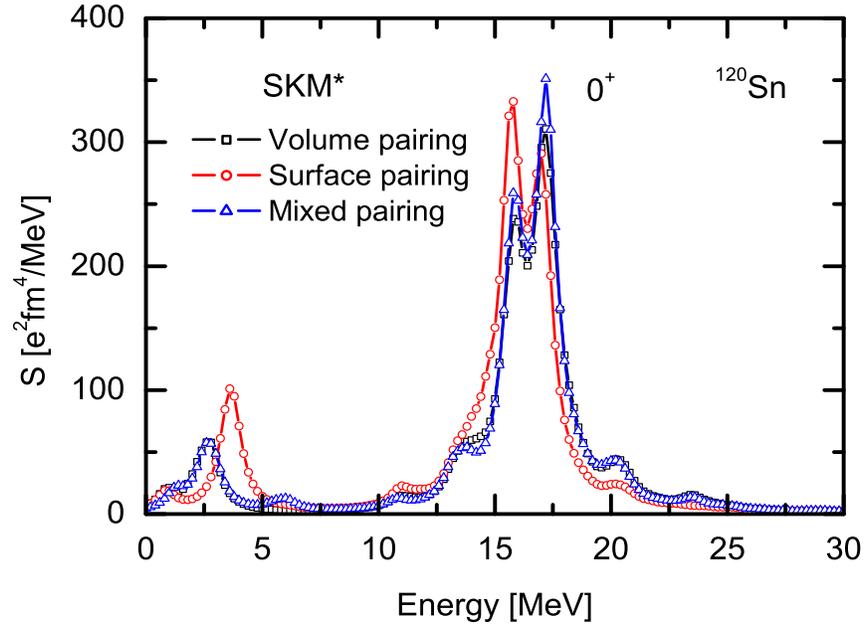}
  \caption{(Color online)The isoscalar $0^{+}$ strength function (averaged with
            Lorentzians having 1 MeV width) in $^{120}$Sn with the Skyrme force SKM*
            for different kinds of pairing forces.}
  \label{fig:fig5}
\end{figure}

\begin{figure}[htbp]
  \centering
  \includegraphics[height=10cm,width=13cm]{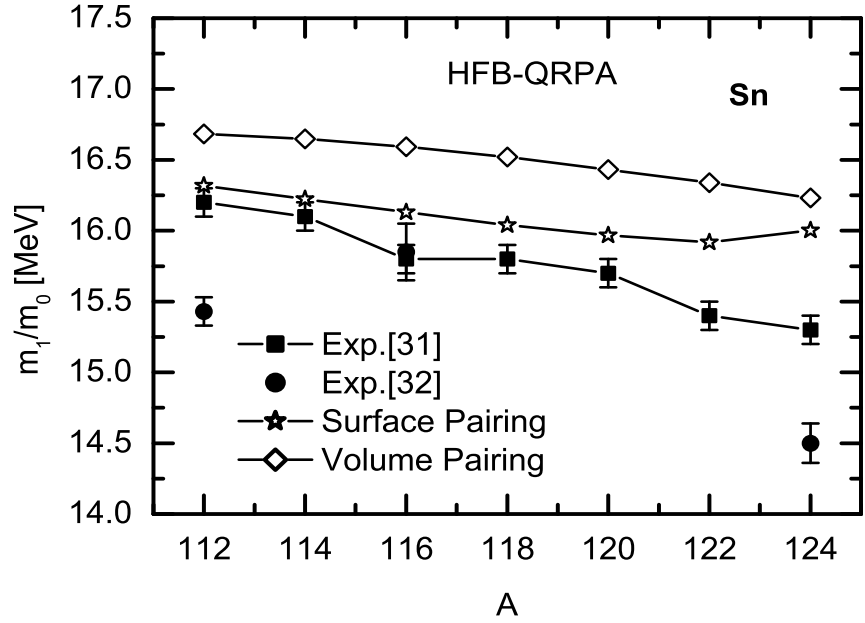}
  \caption{Systematics of the moment ratios $m_{1}/m_{0}$ for the ISGMR
           strength distributions in the tin isotopes using HFB+QRPA with
           the Skyrme force SKM* and volume and surface pairing forces.
           They are evaluated in the energy interval between 10.5 and 20.5 MeV.}
  \label{fig:fig6}
\end{figure}

Already from Fig. \ref{fig:fig3}, it is clear that SKM* performs better than SLy5
in the tin isotopes. Having understood that pairing is also lowering the monopole
energies, as a next step we study the isoscalar $0^{+}$ excitations using HFB+QRPA
with SKM* in the mean field. At the same time, we would also like to discuss in detail
the influence of different kinds of pairing forces on the ISGMR centroid energies.
Fig. \ref{fig:fig5} shows the isoscalar $0^{+}$ strength function (averaged with
Lorentzians having 1 MeV width) in $^{120}$Sn with the Skyrme force SKM* for different
kinds of pairing force. The results with the volume and the
mixed pairing force are almost
the same, while the results with the surface pairing force are different from the others,
especially in the low energy region. This difference comes from the fact that in the
ground state HFB calculations, the energies and occupation numbers of the levels
around the Fermi surface are very similar for volume and mixed pairing, while they are
different for surface pairing. Fig. \ref{fig:fig6} shows the ISGMR centroid energies
in the tin isotopes with SKM* in the mean field and volume and surface pairing forces.
Using the SKM* plus surface pairing force, one can reproduce the
experimental data obtained at RCNP
from $^{112}$Sn to $^{120}$Sn, while SKM* plus the volume pairing force still overestimates
the the centroid energies by about 0.5 MeV. The discrepancy with the
experimental findings from Texas A\& M remains quite significant.

\section{conclusions}

The fully self-consistent QRPA based on the canonical HFB basis
has been introduced by the authors of Ref.~\cite{terasaki2005}. In
the present work, we focus on the same model, having developed
independently the formalism and the computer codes.
Our motivation is the description
of the isoscalar monopole excitations. The main novel aspect of
this paper is the attempt to understand the anomaly in the
experimental ISGMR energies of the Sn isotopes, which cannot be reproduced
using effective forces which do reproduce the experimental energy
in, e.g., $^{208}$Pb; in other words, our goal is to answer the
question raised in Ref.~\cite{piekarewicz2007}, that is, ``why is
tin so soft~?''.

The model uses the zero-range Skyrme force in the
particle-hole channel and the density-dependent pairing force in the particle-particle
channel. The spurious state, caused by the fact that the particle
number symmetry is broken within HFB and restored in QRPA, is
projected out. We investigate the isoscalar 0$^{+}$ strength function
and the ISGMR centroid energies in the tin isotopes. We have also found that the
pairing-rearrangement terms and find they have a negligible
effects (about 0.1\%) on the ISGMR centroid energies in the tin isotopes.
Compared with HF+RPA and HF-BCS-QRPA,
HFB+QRPA makes the theoretical results of ISGMR centroid energies in the tin isotopes
significantly closer to experiment. Pairing improves the absolute values of energy compared
to experiment because of the attractive character of the particle-particle residual
interaction. This amounts to say that the ``softness'' of tin with
respect to monopole excitations is to some extent related to pairing.

Using an appropriate Skyrme force like SKM* and surface pairing force, we
reproduce the experimental ISGMR centroid energies from $^{112}$Sn to $^{120}$Sn.
The pairing energy in nuclear matter, and consequently the effect of pairing on the nuclear
incompressibility $K_{\infty}$, is very small.
Our results imply therefore that, whereas
$^{208}$Pb leads to values of $K_{\infty}$ around 230$\sim$240 MeV, this value is about
10$\%$ smaller when Sn experimental data are used.
Our results do solve partly the puzzle caused by the RCNP experimental results
in the tin isotopes. At the same time,
the 10\% variation of $K_\infty$ from different nuclei points to our still
incomplete understanding of the details of the nuclear effective functionals, in particular
of their density dependence. Also in Fig. 8 of Ref. \cite{vretenar2003}, it is evident that
RMF calculations fit very well the energies of the monopole in Pb and Sn with two values of
$K_{\infty}$ which differ by $\approx$10$\%$. Further investigations on how to constrain
the surface and asymmetry contributions to the final nuclear incompressibility are demanded.

\begin{acknowledgments}

We gratefully acknowledge useful discussions with Jun Terasaki.
This work is partly supported by the Major State Basic Research
Development Program 2007CB815000 as well as the
National Natural Science Foundation of China under Grant No. 10435010,
10775004 and 10221003. We also acknowledge support from the Asia Link
Programme CN/Asia-Link 008 (94791).

\end{acknowledgments}


\clearpage


\begin{table}[ht]
\centering
\caption{The single particle energies (MeV) and occupation number of d3/2 and h11/2 in
         the tin isotopes using the Skyrme force SLy5. In HFB, we use the volume pairing
         force in particle-particle channel.}
\label{tab:table1}
\begin{ruledtabular}
\begin{tabular}{c|cccccc|cccccc}
&\multicolumn{1}{c}{}
&\multicolumn{1}{c}{}
&\multicolumn{1}{c}{}
&\multicolumn{1}{c}{d3/2}
&\multicolumn{1}{c}{}
&\multicolumn{1}{c|}{}
&\multicolumn{1}{c}{}
&\multicolumn{1}{c}{}
&\multicolumn{1}{c}{}
&\multicolumn{1}{c}{h11/2}
&\multicolumn{1}{c}{}
&\multicolumn{1}{c }{}\\
\hline
   & HF & & BCS & & HFB & & HF & & BCS & & HFB & \\
 A & $\epsilon$ & $V^{2}$ & $\epsilon$ & $V^{2}$ & $\epsilon$ & $V^{2}$
   & $\epsilon$ & $V^{2}$ & $\epsilon$ & $V^{2}$ & $\epsilon$ & $V^{2}$  \\
\hline
112 & -8.69 & 0   & -8.63 & 0.18 & -8.33 & 0.21 & -7.14 & 0    & -6.90 & 0.066 & -6.60 & 0.048 \\
114 & -8.74 & 0   & -8.65 & 0.26 & -8.37 & 0.29 & -7.45 & 0    & -7.07 & 0.089 & -6.80 & 0.067 \\
116 & -8.69 & 0   & -8.67 & 0.36 & -8.42 & 0.40 & -7.49 & 0    & -7.23 & 0.12  & -6.98 & 0.095 \\
118 & -8.62 & 0.5 & -8.70 & 0.48 & -8.47 & 0.54 & -7.61 & 0    & -7.35 & 0.16  & -7.11 & 0.14  \\
120 & -8.57 & 1.0 & -8.74 & 0.64 & -8.52 & 0.68 & -7.72 & 0    & -7.46 & 0.21  & -7.22 & 0.20  \\
122 & -8.70 & 1.0 & -8.81 & 0.78 & -8.61 & 0.80 & -7.69 & 0.17 & -7.53 & 0.29  & -7.28 & 0.29  \\
124 & -8.83 & 1.0 & -8.90 & 0.87 & -8.71 & 0.89 & -7.66 & 0.33 & -7.57 & 0.40  & -7.33 & 0.41  \\
\end{tabular}
\end{ruledtabular}
\end{table}

\clearpage


\end{document}